\documentclass[twocolumn]{revtex4-1}
\usepackage{amsmath,amssymb,bm,graphicx,hyperref}
\allowdisplaybreaks[1]

\renewcommand\d{\partial}
\newcommand\R{\mathbb{R}}
\newcommand\Z{\mathbb{Z}}
\newcommand\D{\mathcal{D}}
\newcommand\F{\mathcal{F}}
\newcommand\J{\mathcal{J}}
\newcommand\cri{\mathrm{cri}}
\newcommand\lat{\mathrm{lat}}
\newcommand\G{\mathrm{G}}
\newcommand\M{\mathrm{M}}
\newcommand\W{\mathrm{W}}
\newcommand\XY{\mathrm{XY}}
\DeclareMathOperator\sgn{sgn}
\DeclareMathOperator\tr{tr}

\begin{document}

\title{Boson-fermion duality in four dimensions}

\author{Takuya Furusawa}
\author{Yusuke Nishida}
\affiliation{Department of Physics, Tokyo Institute of Technology,
Ookayama, Meguro, Tokyo 152-8551, Japan}

\date{October 2018}

\begin{abstract}
Dualities provide deep insight into physics by relating two seemingly distinct theories.
Here we propose and elaborate on a novel duality between bosonic and fermionic theories in four spacetime dimensions.
Starting with a Euclidean lattice action consisting of bosonic and fermionic degrees of freedom and integrating out one of them alternatively, we derive a UV duality between a Wilson fermion with self-interactions and an XY model coupled to a compact U(1) gauge field.
We find a continuous phase transition between topological and trivial insulators on the fermion side corresponding to Higgs and confinement phases on the boson side.
The continuum limit of each lattice theory then leads to an IR duality between a free Dirac fermion and a scalar QED with the vacuum angle $\pi$.
The resulting bosonic theory proves to incorporate a scalar boson and dyons as low-energy degrees of freedom and it is their three-body composite that realizes the Dirac fermion of the fermionic theory.
\end{abstract}

\maketitle

\section{Introduction}
Particle statistics is one of the most fundamental outcomes from quantum mechanics, which classifies particles into either boson or fermion.
Although bosons and fermions exhibit quite distinct behaviors, it is possible to transmute particle statistics in three spacetime dimensions (3D) by attaching one magnetic flux quantum to a particle~\cite{Wilczek:1982a,Wilczek:1982b}.
In terms of quantum field theories, the statistical transmutation is implemented by coupling a matter field to a Chern-Simons gauge field~\cite{Polyakov:1988}, which has been successfully employed to account for the fractional quantum Hall effect~\cite{Zhang:1989,Lopez:1991,Halperin:1993}.

Recently, significant progress has been made regarding a web of dualities in 3D.
Here a central position is occupied by the so-called boson-fermion duality, which relates a Wilson-Fisher boson coupled to a U(1) Chern-Simons gauge field at level $1$ and a free massless Dirac fermion~\cite{Karch:2016,Seiberg:2016}.
The boson-fermion duality can be thought of as the statistical transmutation in relativistic critical systems and generates other known dualities such as the particle-vortex dualities for bosons~\cite{Peskin:1978,Dasgupta:1981} and for fermions~\cite{Son:2015,Wang:2016,Metlitski:2016,Mross:2016} as well as new ones~\cite{Karch:2016,Seiberg:2016}.
While the boson-fermion duality was originally proposed on the basis of the conjectured non-Abelian dualities~\cite{Aharony:2016}, it is now explicitly constructed on an array of coupled wires~\cite{Mross:2017} and on a Euclidean lattice~\cite{Chen:2018a}.
In particular, the latter approach has been extended to construct some of the non-Abelian dualities~\cite{Chen:2018b,Jian:arxiv}.

In spite of such great success in 3D, nonsupersymmetric dualities in 4D have been far less explored (see Refs.~\cite{Aitken:2018,Bi:arxiv} for recent proposals).
This work is aimed at paving the way for an unseen web of dualities in 4D by constructing an analog of the boson-fermion duality.
To this end, we employ the lattice construction approach developed in Ref.~\cite{Chen:2018a} and extend it to 4D.
Our starting point is a Euclidean lattice action consisting of the gauged XY model,
\begin{align}
S_\XY[\theta,a] = -\beta\sum_{x,\mu}\cos(\Delta_\mu\theta_x-a_{x,\mu}),
\end{align}
and the gauged Wilson fermion~\cite{Creutz:1983},
\begin{align}\label{eq:wilson}
& S_\W[\chi,a]
= -\sum_{x,\mu}\left(\bar\chi_{x+\mu}\frac{1+\gamma^\mu}{2}e^{ia_{x,\mu}}\chi_x\right. \notag\\
&\left.{} + \bar\chi_x\frac{1-\gamma^\mu}{2}e^{-ia_{x,\mu}}\chi_{x+\mu}\right)
+ (4+M)\sum_x\bar\chi_x\chi_x.
\end{align}
Here $\theta_x\in[-\pi,\pi]$ is an angular variable, $\chi_x$ and $\bar\chi_x$ are four-component Grassmann variables on a site $x\in\Z^4$, and $a_{x,\mu}\in[-\pi,\pi]$ is a compact U(1) gauge field on a link $(x,\mu)$ with $\mu=1,2,3,4$.
The gamma matrices obey $\{\gamma^\mu,\gamma^\nu\}=2\delta^{\mu\nu}$, the lattice constant is set to be unity, and $\beta$ and $M$ are the dimensionless system parameters.
Provided that $\Delta_\mu$ is the forward difference, the action is invariant under the simultaneous gauge transformations of $\theta_x\to\theta_x+\lambda_x$, $\chi_x\to e^{i\lambda_x}\chi_x$, $\bar\chi_x\to\bar\chi_xe^{-i\lambda_x}$, and $a_{x,\mu}\to a_{x,\mu}+\Delta_\mu\lambda_x$.

The partition function then reads
\begin{align}\label{eq:partition}
Z[A] = \int_{-\pi}^\pi\!Da\int_{-\pi}^\pi\!D\theta\int\!D\bar\chi D\chi\,
e^{-S_\XY[\theta,a]-S_\W[\chi,a+A]},
\end{align}
where the Wilson fermion is additionally coupled to an external gauge field $A_{x,\mu}$.
We note that an action for the dynamical gauge field $a_{x,\mu}$ is not included here corresponding to an infinite gauge coupling.
Our strategy is such that the bosonic degrees of freedom are first integrated out to derive a purely fermionic theory and then the fermionic degrees of freedom to derive a purely bosonic theory.
Because the resulting two lattice theories and their continuum limits necessarily share the same correlation functions, they constitute a novel boson-fermion duality in 4D, whose contents are to be elucidated below.

\section{Fermionic theory}
\subsection{Lattice action}
The bosonic degrees of freedom, $\theta_x$ and $a_{x,\mu}$, in Eq.~(\ref{eq:partition}) can be exactly integrated out in a similar way to the 3D case~\cite{Chen:2018a}.
We first change the integration variables as $a_{x,\mu}\to a_{x,\mu}+\Delta_\mu\theta_x$, $\chi_x\to e^{i\theta_x}\chi_x$, and $\bar\chi_x\to\bar\chi_xe^{-i\theta_x}$ so as to eliminate $\theta_x$ from the integrand.
The integration over $\theta_x$ is thus trivial and the partition function for the gauged XY model is brought to
\begin{align}\label{eq:current}
\int_{-\pi}^\pi\!D\theta\,e^{-S_\XY[0,a]}
= \sum_{\{j\}}\prod_{x,\mu}I_{j_{x,\mu}}(\beta)e^{-ij_{x,\mu}a_{x,\mu}},
\end{align}
where the Jacobi-Anger expansion is employed with an integer variable $j_{x,\mu}\in\Z$ on each link representing a current of boson.
On the other hand, the action for the gauged Wilson fermion can be expanded as
\begin{align}\label{eq:expansion}
& e^{-S_\W[\chi,a+A]} = e^{-\sum_x(4+M)\bar\chi_x\chi_x}
\prod_{x,\mu}\sum_{p,q=0}^2\frac1{p!q!} \notag\\
&\times \left(h_{x,\mu}^+[\chi]e^{ia_{x,\mu}+iA_{x,\mu}}\right)^p
\left(h_{x,\mu}^-[\chi]e^{-ia_{x,\mu}-iA_{x,\mu}}\right)^q,
\end{align}
where $h_{x,\mu}^+[\chi]\equiv\bar\chi_{x+\mu}(1+\gamma^\mu)\chi_x/2$ and $h_{x,\mu}^-[\chi]\equiv\bar\chi_x(1-\gamma^\mu)\chi_{x+\mu}/2$ are the forward and backward hopping terms, respectively.
They obey $(h_{x,\mu}^\pm[\chi])^3=0$ because $1\pm\gamma^\mu$ being the matrices of rank $2$ project out two linear combinations of the four Grassmann components.

By combining Eqs.~(\ref{eq:current}) and (\ref{eq:expansion}), the integration over $a_{x,\mu}$ generates $\delta_{j_{x,\mu},p-q}$ on each link, which equalizes the currents of boson and fermion indicating that they are bound together at the infinite gauge coupling.
After the summation over $j_{x,\mu}$, we thus obtain
\begin{align}
Z[A] &= \int\!D\bar\chi D\chi\,e^{-\sum_x(4+M)\bar\chi_x\chi_x}
\prod_{x,\mu}\sum_{p,q=0}^2\frac{I_{p-q}(\beta)}{p!q!} \notag\\
&\quad \times \left(h_{x,\mu}^+[\chi]e^{iA_{x,\mu}}\right)^p
\left(h_{x,\mu}^-[\chi]e^{-iA_{x,\mu}}\right)^q.
\end{align}
Finally, by rescaling the fermion fields as $\chi_x=\sqrt{I_0(\beta)/I_1(\beta)}\,\psi_x$ and $\bar\chi_x=\sqrt{I_0(\beta)/I_1(\beta)}\,\bar\psi_x$ and reexponentiating the hopping terms, the partition function up to overall factors is provided by
\begin{align}\label{eq:fermion_lattice}
Z[A] \propto \int\!D\bar\psi D\psi\,e^{-S'_\W[\psi,A]-S_\mathrm{int}[\psi,A]},
\end{align}
where $S'_\W[\psi,A]$ is the action for the Wilson fermion in Eq.~(\ref{eq:wilson}) with its bare mass $M$ replaced by $M'\equiv(4+M)I_0(\beta)/I_1(\beta)-4$.
Six self-interaction terms are now included in
\begin{align}\label{eq:interaction}
& S_\mathrm{int}[\psi,A] \notag\\
&= \sum_{x,\mu}\sum_{p+q\geq2}\frac{U_{p,q}}{p!q!}
\left(h_{x,\mu}^+[\psi]e^{iA_{x,\mu}}\right)^p\left(h_{x,\mu}^-[\psi]e^{-iA_{x,\mu}}\right)^q
\end{align}
and the corresponding $(p{\,+\,}q)$-body coupling constants read $U_{1,1}=1-[I_0(\beta)/I_1(\beta)]^2$, $U_{2,0}=U_{0,2}=1-I_0(\beta)I_2(\beta)/[I_1(\beta)]^2$, $U_{2,1}=U_{1,2}=-U_{1,1}-U_{2,0}$, and $U_{2,2}=2U_{1,1}+2U_{2,0}+(U_{1,1})^2+(U_{2,0})^2$.
Their magnitudes are controlled by $1/\beta$ originally playing a role of temperature in the gauged XY model.
The resulting fermionic theory in Eq.~(\ref{eq:fermion_lattice}) is the self-interacting Wilson fermion, which constitutes the fermion side of our boson-fermion duality at the lattice level.

\subsection{Phase diagram}
The fermionic theory being obtained, we then elucidate its phase structure in the parameter space of $(M,1/\beta)$ with $A_{x,\mu}$ temporarily turned off.
In particular, when $1/\beta=+0$, all the coupling constants in Eq.~(\ref{eq:interaction}) vanish so that the Wilson fermion becomes free with its mass provided by $M'|_{1/\beta=+0}=M$.
Therefore, the mass gap closes at $M=0$, which is a critical point separating two gapped phases.
One gapped phase on the side of $M>0$ is a trivial insulator because it is adiabatically connected to the atomic limit $M\to+\infty$, while the other gapped phase on the side of $M<0$ is known to be a celebrated topological insulator~\cite{Qi:2008}.
They cannot be adiabatically connected to each other as long as the time-reversal symmetry is respected~\cite{Fu:2007,Moore:2007,Roy:2009}.
We also note that other critical points exist at $M=-2,-4,-6,-8$, where doublers become massless, but they are not considered here.

\begin{figure}[b]
\includegraphics[width=0.9\columnwidth,clip]{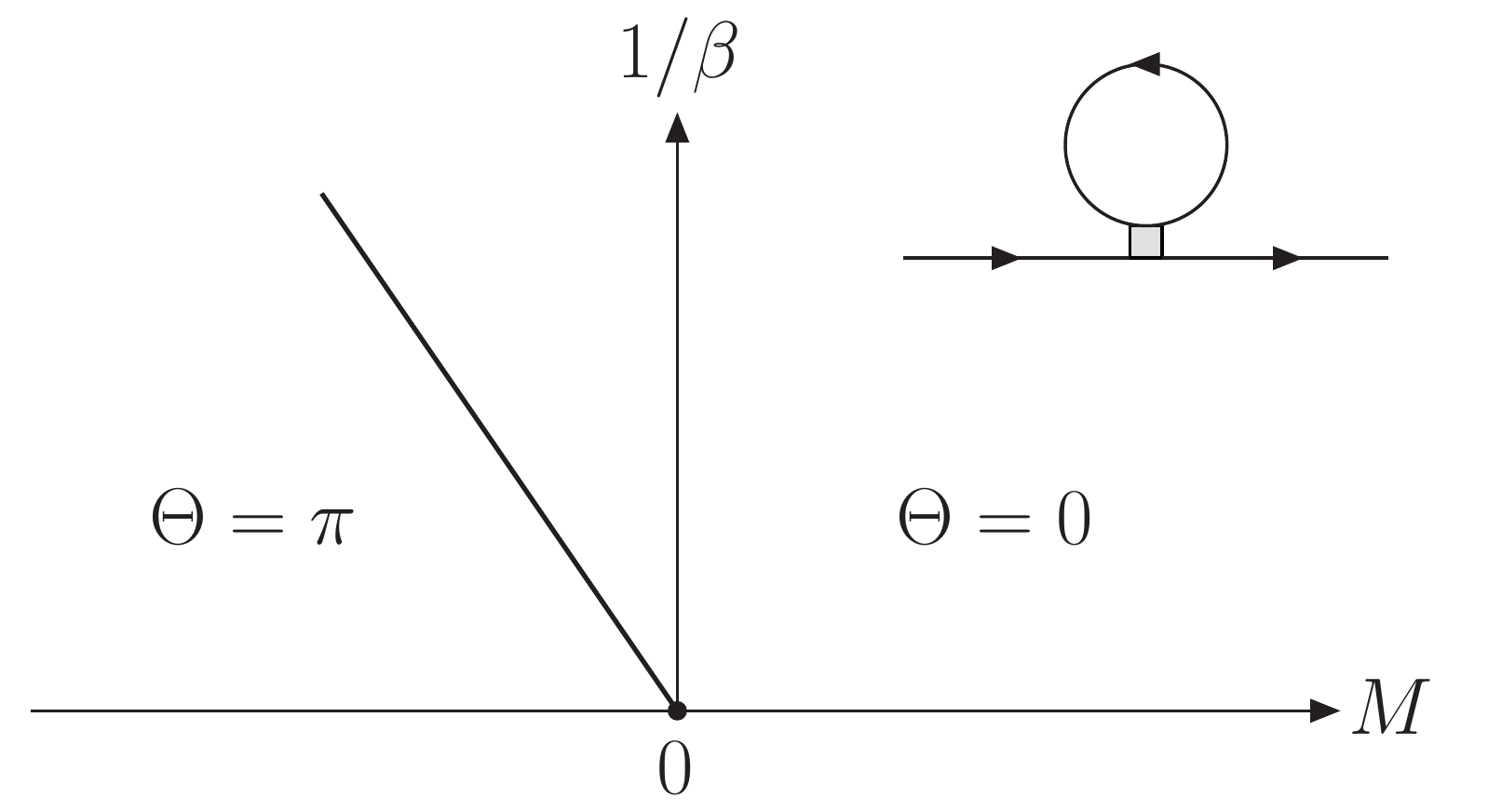}
\caption{\label{fig:phase_diagram}
Phase diagram for Eq.~(\ref{eq:partition}) in the vicinity of $(M,1/\beta)=(0,0)$, where the trivial ($\Theta=0$) and topological ($\Theta=\pi$) insulators are separated by the critical line along $M=-1.94/\beta+O(1/\beta^2)$.
The inset depicts a type of Feynman diagrams contributing to the self-energy at $O(1/\beta)$.}
\end{figure}

How the critical point separating the trivial and topological insulators extends toward $1/\beta>0$ can be determined perturbatively as long as $1/\beta\ll1$.
Because $U_{1,1}=-1/\beta+O(1/\beta^2)$, $U_{2,0}=1/\beta+O(1/\beta^2)$, but $U_{2,1},U_{2,2}=O(1/\beta^2)$, only the two-body coupling constants need to be kept to the leading order in $1/\beta$.
The Wilson fermion propagator is provided by $G(k)=1/[\sum_\mu(i\gamma^\mu\sin k_\mu-\cos k_\mu)+4+M'+\Sigma(k)]$, where the self-energy from a standard diagrammatic calculation (see the inset of Fig.~\ref{fig:phase_diagram}) is found to be
\begin{align}
& \Sigma(k) = \frac1\beta\sum_\mu(i\gamma^\mu\sin k_\mu-\cos k_\mu)
\int_{-\pi}^\pi\!\frac{d^4l}{(2\pi)^4} \notag\\
&\times \frac{\sin^2l_\mu - \cos l_\mu\left[\sum_\nu(1-\cos l_\nu)+M'\right]}
{\sum_\nu\sin^2l_\nu + \left[\sum_\nu(1-\cos l_\nu)+M'\right]^2} + O(1/\beta^2).
\end{align}
The mass gap now closes at $M'+\Sigma(0)=0$, which leads to $M=-1.94/\beta+O(1/\beta^2)$ as depicted in Fig.~\ref{fig:phase_diagram}.
Here the upper right side of the critical line corresponds to the trivial insulator with $M'+\Sigma(0)>0$, while its lower left side corresponds to the topological insulator with $M'+\Sigma(0)<0$.

\subsection{Continuum action}
The continuous phase transition between the trivial and topological insulators allows us to take the continuum limit of the fermionic theory so far defined on the lattice.
Because all the self-interactions in Eq.~(\ref{eq:interaction}) are irrelevant about the Gaussian fixed point corresponding to a free massless Dirac fermion, the fermionic action in Eq.~(\ref{eq:fermion_lattice}) is reduced to
\begin{align}\label{eq:fermion_continuum}
S_\mathrm{F}[\psi,A] = \int\!d^4x\,\bar\psi[\gamma^\mu(\d_\mu-iA_\mu)+m]\psi.
\end{align}
Here the continuum limit $a_\lat\to0$ is taken with the physical mass of $m=[M'+\Sigma(0)]/a_\lat$ kept constant and summations over repeated indices are implicitly assumed in continuum actions.

Because the resulting free Dirac fermion in Eq.~(\ref{eq:fermion_continuum}) is massive when $m\neq0$, it can be further integrated out leading to
\begin{align}\label{eq:effective}
-\ln Z[A] \to S_\M[A]
+ \frac{i\Theta}{32\pi^2}\int\!d^4x\,\epsilon^{\mu\nu\sigma\tau}F_{\mu\nu}[A]F_{\sigma\tau}[A],
\end{align}
where $S_\M[A]\propto\int\!d^4x(F_{\mu\nu}[A])^2$ with $F_{\mu\nu}[A]=\d_\mu A_\nu-\d_\nu A_\mu$ is the Maxwell action for the external gauge field $A_\mu$ and the vacuum angle is provided by $\Theta=\pi[1-\sgn(m)]/2\pmod{2\pi}$~\cite{Qi:2008,Hosur:2010}.
Therefore, the low-energy effective action is found to be the usual electrodynamics with $\Theta=0$ for the trivial insulator $m>0$, but the axion electrodynamics with $\Theta=\pi$ for the topological insulator $m<0$~\cite{Wilczek:1987,Qi:2008}, as indicated in Fig.~\ref{fig:phase_diagram}.
This completes the fermion side of our boson-fermion duality in the continuum limit.

\section{Bosonic theory}
\subsection{Lattice action}
Turning to the boson side of our boson-fermion duality, the fermionic degrees of freedom, $\chi_x$ and $\bar\chi_x$, in Eq.~(\ref{eq:partition}) can be formally integrated out leading to
\begin{align}\label{eq:boson_lattice}
Z[A] = \int_{-\pi}^\pi\!Da\int_{-\pi}^\pi\!D\theta\,e^{-S_\XY[\theta,a]-S_\G[a+A]},
\end{align}
where $S_\G[a]=-\tr\ln(\d^2S_\W[\chi,a]/\d\bar\chi_x\d\chi_y)$ is from the Wilson fermion determinant.
Because the action for the dynamical gauge field is now generated, the resulting bosonic theory in Eq.~(\ref{eq:boson_lattice}) can be thought of as a variant of the Abelian Higgs model, which proves to be dual to the self-interacting Wilson fermion in Eq.~(\ref{eq:fermion_lattice}) at the lattice level.
They share the same partition function and thus must share the same phase diagram depicted in Fig.~\ref{fig:phase_diagram}, consisting of the two gapped phases separated by the critical line along $1/\beta=-0.516M+O(M^2)\equiv1/\beta_\cri$ on the side of $M<0$.

In order to understand the two gapped phases in terms of the bosonic theory, we recall that phases of the Abelian Higgs model are typically classified into Coulomb, Higgs, and confinement phases~\cite{Fradkin:1979,Borgs:1987}.
Although the Coulomb phase is ineligible for our phase diagram because the gauge field therein is massless, it acquires a mass gap in the other two phases~\cite{Osterwalder:1978}.
It is then reasonable to identify the gapped phase at $1/\beta<1/\beta_\cri$ as the Higgs phase because XY spins tend to order at lower temperature, while the confinement phase is left for the other gapped phase at $1/\beta>1/\beta_\cri$.
The latter identification is also reasonable because it is adiabatically connected to the limit $M\to+\infty$, where the action for the dynamical gauge field is reduced to the simple plaquette action, $S_\G[a]\to-1/(2M^4)\sum_{x,\mu,\nu}\cos(\Delta_\mu a_{x,\nu}-\Delta_\nu a_{x,\mu})$, with an infinite gauge coupling.

\begin{table}[b]
\caption{\label{tab:correspondence}
Understanding of the two gapped phases in Fig.~\ref{fig:phase_diagram} in terms of the fermionic theory in Eqs.~(\ref{eq:fermion_lattice}) and (\ref{eq:fermion_continuum}) versus the bosonic theory in Eqs.~(\ref{eq:boson_lattice}) and (\ref{eq:boson_continuum}).}
\begin{ruledtabular}
\begin{tabular}{c|cc}
& $\Theta=\pi$ & $\Theta=0$ \\[2pt]\hline
Fermion & Topological \ ($m<0$) & Trivial \ ($m>0$) \\
Boson & Higgs \ ($r<0$) & Confinement \ ($r>0$)
\end{tabular}
\end{ruledtabular}
\end{table}

The above understanding and the correspondence to the fermion side are summarized in Table~\ref{tab:correspondence}, which is to be further confirmed below by reproducing the low-energy effective action in Eq.~(\ref{eq:effective}) from the boson side.
We note that no order parameter can distinguish the Higgs and confinement phases with the fundamental charge~\cite{Fradkin:1979}, which is indeed consistent with the fact that no order parameter can distinguish the topological and trivial insulators on the fermion side.
Unlike in the typical Abelian Higgs model, our Higgs and confinement phases cannot be adiabatically connected to each other as long as the time-reversal symmetry is respected because of the distinct vacuum angles $\Theta$.

\subsection{Continuum action}
The continuous phase transition between the Higgs and confinement phases is controlled by $1/\beta$ on the side of $M<0$, where it is possible in principle to take the continuum limit of the bosonic theory so far defined on the lattice.
Although it is intractable in practice because the system is strongly coupled about the critical point, we proceed by assuming that the continuum limit may be considered separately for $S_\XY[\theta,a]$ and $S_\G[a]$ in Eq.~(\ref{eq:boson_lattice}).
First, regarding the gauged XY model, it is convenient to think of it as a limiting case of the gauged complex scalar model, $S_\XY[\theta,a]=-\sum_{x,\mu}(\phi_{x+\mu}^*e^{ia_{x,\mu}}\phi_x+\phi_x^*e^{-ia_{x,\mu}}\phi_{x+\mu})+u\sum_x(|\phi_x|^2-\beta/2)^2|_{u\to+\infty}$, with $\phi_x\sim e^{i\theta_x}$.
Its continuum limit is then provided by the Ginzburg-Landau action, $S_\XY[\theta,a]\to\int\!d^4x[|(\d_\mu-ia_\mu)\phi|^2+r|\phi|^2+u|\phi|^4]$, where the continuous phase transition is controlled by $r$ so that $r\lessgtr0$ with $r=0$ defined as the critical point corresponds to $1/\beta\lessgtr1/\beta_\cri$.

On the other hand, regarding the pure gauge action of $S_\G[a]$, we recall that it is generated by integrating out $\chi_x$ and $\bar\chi_x$ in Eq.~(\ref{eq:partition}).
Because the mass of the bare Wilson fermion is negative on the side of $M<0$, $S_\G[a]$ is reduced to the Maxwell action with the vacuum angle $\pi$, $S_\G[a]\to S_\M[a]+i/(32\pi)\int\!d^4x\,\epsilon^{\mu\nu\sigma\tau}F_{\mu\nu}[a]F_{\sigma\tau}[a]$, in the continuum limit [see Eq.~(\ref{eq:effective})].
Therefore, the above consideration leads us to propose that the bosonic action in Eq.~(\ref{eq:boson_lattice}) turns into the form of
\begin{align}\label{eq:boson_continuum}
& S_\mathrm{B}[\phi,a,A]
= \int\!d^4x\!\left[|(\d_\mu-ia_\mu)\phi|^2 + r|\phi|^2 + u|\phi|^4\right] \notag\\
&+ S_\M[a+A] + \frac{i}{32\pi}\int\!d^4x\,\epsilon^{\mu\nu\sigma\tau}F_{\mu\nu}[a+A]F_{\sigma\tau}[a+A].
\end{align}
This is none other than the scalar QED with the vacuum angle $\pi$, which proves to be dual to the free Dirac fermion in Eq.~(\ref{eq:fermion_continuum}) with $r\lessgtr0$ corresponding to $m\lessgtr0$.
We note that the dynamical gauge field $a_\mu$ needs to be understood as compact allowing for magnetic monopoles and thus the confinement phase.

The resulting bosonic theory in Eq.~(\ref{eq:boson_continuum}) must be consistent with the low-energy effective action obtained from the fermion side.
First, in the Higgs phase with $r<0$, the scalar field condenses and thus generates a mass term $\propto(a_\mu)^2$ written in the unitary gauge.
Then, by integrating out the massive gauge field with higher derivative terms neglected, Eq.~(\ref{eq:effective}) is reproduced with $\Theta=\pi$.
On the other hand, in the confinement phase with $r>0$, the scalar field is massive and is thus integrated out leaving behind $S_\M[a]$.
When $S_\M[a]$ is absent corresponding to the limit of $r\to+\infty$, no effective action for $A_\mu$ is left after the integration over $a_\mu$.
Then, by turning on $S_\M[a]$ adiabatically within the confinement phase, the Maxwell action for $A_\mu$ may be generated with $\Theta=0$ protected by the time-reversal symmetry so as to reproduce Eq.~(\ref{eq:effective}).
Therefore, the topological and trivial insulators on the fermion side are consistently realized by the Higgs and confinement phases on the boson side, as shown in Table~\ref{tab:correspondence}.

\subsection{Dyons}
In order to gain further insight into the confinement phase, we need to handle the compact gauge field, which is conveniently formulated on a Euclidean lattice again.
In the Villain representation, the pure gauge action resulting from Eq.~(\ref{eq:boson_continuum}) at $r\to+\infty$ reads
\begin{align}\label{eq:gauge}
& S[a,A] = \frac1{4e^2}\sum_{x,\mu,\nu}(\F_{x,\mu\nu}[a+A])^2 \notag\\
&+ \frac{i\Xi}{32\pi^2}\sum_{x,\mu,\nu,\sigma,\tau}\epsilon^{\mu\nu\sigma\tau}\F_{x,\mu\nu}[a+A]\F_{x,\sigma\tau}[a+A],
\end{align}
where $\Xi=\pi$ in our case and $\F_{x,\mu\nu}[a]=\Delta_\mu a_{x,\nu}-\Delta_\nu a_{x,\mu}+2\pi n_{x,\mu\nu}$ is the field strength tensor with an integer variable $n_{x,\mu\nu}\in\Z$ on each plaquette~\cite{Cardy:1982a,Cardy:1982b}.
The monopole current is provided by $\J_{x,\mu}=1/(4\pi)\sum_{\nu,\sigma,\tau}\epsilon^{\mu\nu\sigma\tau}\Delta_\nu\F_{x,\sigma\tau}[a]=1/2\sum_{\nu,\sigma,\tau}\epsilon^{\mu\nu\sigma\tau}\Delta_\nu n_{x,\sigma\tau}\in\Z$ and satisfies the conservation law, $\sum_\mu\Delta_\mu\J_{x,\mu}=0$, automatically.
The presence of such magnetic monopoles can be made explicit by bringing Eq.~(\ref{eq:gauge}) to its dual representation~\cite{Peskin:1978,Herbut:2007},
\begin{align}\label{eq:dual}
S'[\zeta,b] &= -\beta'\sum_{x,\mu}\cos(\nabla_{\!\mu}\zeta_x-b_{x,\mu})
+ \frac1{4e'^2}\sum_{x,\mu,\nu}(f_{x,\mu\nu}[b])^2 \notag\\
&\quad + \frac{i\Xi'}{32\pi^2}\sum_{x,\mu,\nu,\sigma,\tau}\epsilon^{\mu\nu\sigma\tau}f_{x,\mu\nu}[b]f_{x,\sigma\tau}[b],
\end{align}
where $\nabla_{\!\mu}$ is the backward difference, $\zeta_x\in[-\pi,\pi]$ is an angular variable, and $b_{x,\mu}\in\R$ is a noncompact U(1) gauge field on a link $(x,-\mu)$ with its field strength tensor provided by $f_{x,\mu\nu}[b]=\nabla_{\!\mu}b_{x,\nu}-\nabla_{\!\nu}b_{x,\mu}$.
The dimensionless system parameters are found to be $1/\beta'=+0$, $e'^2=(2\pi/e)^2+(\Xi e/2\pi)^2$, and $\Xi'=-\Xi e^2/e'^2$, as detailed in the Appendix.

The above two lattice actions are related via the electromagnetic duality and share the same partition function up to overall factors.
Corresponding to the Coulomb and confinement phases of the pure gauge theory in Eq.~(\ref{eq:gauge}), the noncompact Abelian Higgs model in Eq.~(\ref{eq:dual}) is known to exhibit the Coulomb and Higgs phases, at least when $\Xi=0$~\cite{Borgs:1987}.
In particular, in the Higgs phase with a suitable gauge fixing, the scalar field $\D_x\sim e^{i\zeta_x}$ condenses and thus generates a mass gap for the gauge field~\cite{Kennedy:1985,Kennedy:1986}.
Because $f_{x,\mu\nu}[b]$ is introduced as a Hubbard-Stratonovich field coupled to $\frac{\Xi e}{2\pi}(\frac1e\F_{x,\mu\nu}[a+A])+\frac{2\pi}{e}(-\frac{i}{2e}\sum_{\sigma,\tau}\epsilon^{\mu\nu\sigma\tau}\F_{x,\sigma\tau}[a+A])$ [see Eqs.~(\ref{eq:EoM}) and (\ref{eq:CoV}) in the Appendix], the electrically charged $\D_x$ under $b_{x,\mu}$ actually carries both electric and magnetic charges of $\Xi e/2\pi$ and $2\pi/e$, respectively, under $a_{x,\mu}$ and $A_{x,\mu}$.
Therefore, it is a magnetic monopole when $\Xi=0$ but becomes a dyon in our case with $\Xi=\pi$ because of the Witten effect~\cite{Witten:1979}, whose condensation leads to the confinement of electric charges.

\begin{table}[t]
\caption{\label{tab:charge}
Charges and particle statistics of a scalar boson $\phi$, a dyon $\D$, its time-reversal partner $\tilde\D$, and their composites in Eq.~(\ref{eq:boson_continuum}) versus a Dirac fermion $\psi$ in Eq.~(\ref{eq:fermion_continuum}).
Shown are the electric ($q_e,Q_e$) and magnetic ($q_m,Q_m$) charges under the dynamical $a_\mu$ (lower case) and external $A_\mu$ (upper case) gauge fields, while the particle statistics is indicated by B (boson) or F (fermion).}
\begin{ruledtabular}
\begin{tabular}{c|ccccc}
& $q_e$ & $q_m$ & $Q_e$ & $Q_m$ & Stat. \\[2pt]\hline
$\phi$ & 1 & 0 & 0 & 0 & B \\
$\D$ & 1/2 & 1 & 1/2 & 1 & B \\
$\tilde\D$ & 1/2 & $-1$ & 1/2 & $-1$ & B \\
$\D\tilde\D$ & 1 & 0 & 1 & 0 & F \\
$\phi^*\D\tilde\D$ & 0 & 0 & 1 & 0 & F \\
$\psi$ & 0 & 0 & 1 & 0 & F
\end{tabular}
\end{ruledtabular}
\end{table}

We thus find that the bosonic theory in Eq.~(\ref{eq:boson_continuum}) incorporates a dyon $\D$ as an elementary degree of freedom relevant to low-energy physics.
It is a boson carrying the electric and magnetic charges of $(q_e,q_m)=(1/2,1)$ under $a_\mu$ and $(Q_e,Q_m)=(1/2,1)$ under $A_\mu$ in units of $(e,2\pi/e)$.
Its time-reversal partner denoted by $\tilde\D$ is also a boson and carries $(q_e,q_m)=(Q_e,Q_m)=(1/2,-1)$, while a pair of $\D\tilde\D$ carrying $q_e=Q_e=1$ is transmuted into a fermion because of the charge-monopole statistical interaction~\cite{Goldhaber:1976,Metlitski:2013,Wang:2014,Metlitski:2015}.
Then, combined with the scalar boson $\phi$ carrying $q_e=1$, a three-body composite of $\phi^*\D\tilde\D$ remains a fermion but carries $Q_e=1$ only just as the Dirac fermion $\psi$ in Eq.~(\ref{eq:fermion_continuum}).
Therefore, as compared in Table~\ref{tab:charge}, the Dirac fermion on the fermion side is consistently realized by the three-body composite of the scalar boson and dyons on the boson side.
All the correlation functions with respect to them should match each other between the bosonic and fermionic theories, which constitutes our boson-fermion duality in the continuum limit.

\section{Summary}
We have proposed and elaborated on the novel boson-fermion duality in 4D.
Starting with the Euclidean lattice action in Eq.~(\ref{eq:partition}) and integrating out the bosonic or fermionic degrees of freedom alternatively, we derived the UV duality between the self-interacting Wilson fermion in Eq.~(\ref{eq:fermion_lattice}) and the variant Abelian Higgs model in Eq.~(\ref{eq:boson_lattice}).
The continuum limit of each lattice theory then led to the IR duality between Eqs.~(\ref{eq:fermion_continuum}) and (\ref{eq:boson_continuum}), which reads
$$\text{free Dirac fermion $\,\leftrightarrow\,$ scalar QED with vacuum angle $\pi$}.$$
The resulting bosonic theory incorporates a scalar boson and dyons as low-energy degrees of freedom and they condense in the Higgs and confinement phases, respectively.
These two gapped phases with the distinct vacuum angles $\Theta$ in Eq.~(\ref{eq:effective}) are separated by a new type of continuous phase transition, i.e., Higgs-confinement criticality, which is dual to the topological phase transition on the fermion side.
It is the three-body composite of the scalar boson and dyons on the boson side that realizes the Dirac fermion of the fermionic theory.

Regarding future works, it is important to further elucidate and confirm our boson-fermion duality, in particular, on the strongly coupled boson side.
To this end, it may be useful to employ complementary approaches such as the operator-based coupled-wire construction~\cite{Mross:2016,Mross:2017} as well as numerical simulations being successful in 3D~\cite{Senthil:arxiv}.
The extension to non-Abelian dualities is currently under way.
Last but not least, whether our boson-fermion duality generates a web of dualities in 4D is indeed worth exploring.

\acknowledgments
The authors thank Yohei Fuji, Akira Furusaki, Shigeki Onoda, Shinsei Ryu, Yuya Tanizaki, and Keisuke Totsuka for the valuable discussions.
This work was supported by JSPS KAKENHI Grants No.~JP15K17727 and No.~JP15H05855.

\onecolumngrid\appendix
\section{Derivation of Eq.~(\ref{eq:dual}) from Eq.~(\ref{eq:gauge})}
Here we present how to derive Eq.~(\ref{eq:dual}) from Eq.~(\ref{eq:gauge}) via the electromagnetic duality based on Refs.~\cite{Peskin:1978,Herbut:2007}.
Starting with the partition function for the pure gauge theory in Eq.~(\ref{eq:gauge}),
\begin{align}
Z = \sum_{\{n\}}\int_{-\pi}^\pi\!Da\,
\exp\!\left[-\frac1{4\pi}\sum_x\sum_{\mu<\nu}\sum_{\sigma<\tau}\F_{x,\mu\nu}[a+A]
\left(\frac{2\pi}{e^2}\delta^{\mu\sigma}\delta^{\nu\tau}
+ \frac{i\Xi}{2\pi}\epsilon^{\mu\nu\sigma\tau}\right)\F_{x,\sigma\tau}[a+A]\right],
\end{align}
the Hubbard-Stratonovich transformation leads to
\begin{align}\label{eq:transformation}
Z \propto \sum_{\{n\}}\int_{-\pi}^\pi\!Da\int_{-\infty}^\infty\!Dg\,
\exp\!\left[-\frac1{4\pi}\sum_x\sum_{\mu<\nu}\sum_{\sigma<\tau}g_{x,\mu\nu}
\left(\frac{2\pi}{e^2}\delta^{\mu\sigma}\delta^{\nu\tau}
+ \frac{i\Xi}{2\pi}\epsilon^{\mu\nu\sigma\tau}\right)^{-1}g_{x,\sigma\tau}
- \frac{i}{2\pi}\sum_x\sum_{\mu<\nu}\F_{x,\mu\nu}[a+A]g_{x,\mu\nu}\right].
\end{align}
This identity is readily confirmed with the equation of motion,
\begin{align}\label{eq:EoM}
g_{x,\mu\nu}\bigg|_\mathrm{EoM}
= -i\sum_{\sigma<\tau}\left(\frac{2\pi}{e^2}\delta^{\mu\sigma}\delta^{\nu\tau}
+ \frac{i\Xi}{2\pi}\epsilon^{\mu\nu\sigma\tau}\right)\F_{x,\sigma\tau}[a+A],
\end{align}
and the matrix inversion under $\mu<\nu$ and $\sigma<\tau$ reads
\begin{align}
\left(\frac{2\pi}{e^2}\delta^{\mu\sigma}\delta^{\nu\tau}
+ \frac{i\Xi}{2\pi}\epsilon^{\mu\nu\sigma\tau}\right)^{-1}
&= \frac{2\pi}{(2\pi/e)^2+(\Xi e/2\pi)^2}\delta^{\mu\sigma}\delta^{\nu\tau}
- \frac{i\Xi}{2\pi}\frac{e^2}{(2\pi/e)^2+(\Xi e/2\pi)^2}\epsilon^{\mu\nu\sigma\tau} \\
&\equiv \frac{2\pi}{e'^2}\delta^{\mu\sigma}\delta^{\nu\tau}
+ \frac{i\Xi'}{2\pi}\epsilon^{\mu\nu\sigma\tau}.
\end{align}
Then, with the understanding of $g_{x,\mu\nu}=-g_{x,\nu\mu}$ for $\mu>\nu$, the integration over $a_{x,\mu}$ in Eq.~(\ref{eq:transformation}) generates a Kronecker delta imposing $\sum_\nu\!\nabla_{\!\nu}g_{x,\mu\nu}=0$ on each link, which is automatically satisfied by changing the variable from $g_{x,\mu\nu}$ to $b_{x,\mu}\in\R$ so that
\begin{align}\label{eq:CoV}
g_{x,\mu\nu} = \sum_{\sigma,\tau}\epsilon^{\mu\nu\sigma\tau}\nabla_{\!\sigma}b_{x,\tau}
= \frac12\sum_{\sigma,\tau}\epsilon^{\mu\nu\sigma\tau}f_{x,\sigma\tau}[b].
\end{align}
Therefore, the partition function is now provided by
\begin{align}\label{eq:Z}
Z = \int_{-\infty}^\infty\!Db\,z[b]
\exp\!\left[-\frac1{4\pi}\sum_x\sum_{\mu<\nu}\sum_{\sigma<\tau}f_{x,\mu\nu}[b]
\left(\frac{2\pi}{e'^2}\delta^{\mu\sigma}\delta^{\nu\tau}
+ \frac{i\Xi'}{2\pi}\epsilon^{\mu\nu\sigma\tau}\right)f_{x,\sigma\tau}[b]\right]
\end{align}
with
\begin{align}
z[b] = \sum_{\{n\}}\exp\!\left[-i\sum_x\sum_{\mu,\nu}\sum_{\sigma<\tau}
\epsilon^{\mu\nu\sigma\tau}b_{x,\mu}\Delta_\nu n_{x,\sigma\tau}\right]
= \sum_{\{\J\}}\int_{-\pi}^\pi\!D\zeta\,
\exp\!\left[-i\sum_{x,\mu}\zeta_x\,\Delta_\mu\J_{x,\mu} - i\sum_{x,\mu}b_{x,\mu}\J_{x,\mu}\right].
\end{align}
Here the variable is changed again from $n_{x,\mu\nu}$ to $\J_{x,\mu}=\sum_\nu\sum_{\sigma<\tau}\epsilon^{\mu\nu\sigma\tau}\Delta_\nu n_{x,\sigma\tau}\in\Z$, which must be constrained by $\sum_\mu\Delta_\mu\J_{x,\mu}=0$ on each site as ensured by the integration over $\zeta_x$.
Finally, the Jacobi-Anger expansion with the asymptotic form of the modified Bessel function,
\begin{align}
e^{\beta\cos\theta} = \sum_{N\in\Z}I_N(\beta)e^{iN\theta}
\underset{\beta\to+\infty}\to \frac{e^{\beta}}{\sqrt{2\pi\beta}}\sum_{N\in\Z}e^{-\frac{N^2}{2\beta}+iN\theta},
\end{align}
is employed on each link leading to
\begin{align}
z[b] &= \lim_{\beta'\to+\infty}\sum_{\{\J\}}\int_{-\pi}^\pi\!D\zeta\,
\exp\!\left[-\frac1{2\beta'}\sum_{x,\mu}(\J_{x,\mu})^2
+ i\sum_{x,\mu}\J_{x,\mu}\nabla_{\!\mu}\zeta_x - i\sum_{x,\mu}\J_{x,\mu}b_{x,\mu}\right] \\\label{eq:z}
&\propto \lim_{\beta'\to+\infty}\int_{-\pi}^\pi\!D\zeta\,
\exp\!\left[\beta'\sum_{x,\mu}\cos(\nabla_{\!\mu}\zeta_x-b_{x,\mu})\right].
\end{align}
By combining Eqs.~(\ref{eq:Z}) and (\ref{eq:z}), we arrive at the partition function for the noncompact Abelian Higgs model in Eq.~(\ref{eq:dual}).

\twocolumngrid

\end{document}